\documentclass[9pt,journal,compsoc]{IEEEtran}
\usepackage{graphicx}
\usepackage{soul}
\usepackage{xspace}
\usepackage{amsmath, mathtools}
\usepackage{xcolor}
\usepackage{cleveref}

\crefformat{section}{\S#2#1#3} 
\crefformat{subsection}{\S#2#1#3}
\crefformat{subsubsection}{\S#2#1#3}

\ifCLASSOPTIONcompsoc
  \usepackage[nocompress]{cite}
\else
  \usepackage{cite}
\fi
\ifCLASSINFOpdf
\else
\fi
\newcommand{\mypar}[1]{\noindent \textbf{#1:}}

\newcommand{\fixme}[1]{\textbf{FIXME: [#1]}}

\newcommand{\eg}{\textit{e.g.}}

\newcommand{\ignore}[1]{}

\newcommand{\revisionhighlight}[1]{#1}

\newcommand\douppercase[1]{\ifnum\ifhmode\spacefactor\else2000\fi>1000 \uppercase{#1}\else#1\fi}

\newcommand{\ramp}{RAMP\xspace}

\newcommand{\pbdue}{{p_{b,due}}}
\newcommand{\pbnde}{{p_{b,nde}}}
\newcommand{\pldue}{{p_{lb,due}}}
\newcommand{\plnde}{{p_{lb,nde}}}
\newcommand{\pcdue}{{p_{c,due}}}
\newcommand{\pcnde}{{p_{c,nde}}}

\begin{document}
%
\title{The Case for Replication-Aware Memory-Error Protection in Disaggregated Memory}
%
%
%
%

\author{Haris Volos\\University of Cyprus\\\texttt{haris.volos@cs.ucy.ac.cy}
\thanks{
 Manuscript received May 30, 2021; revised July 5, 2021.
}}

%
%

\markboth{Computer Architecture Letters}%
{Shell \MakeLowercase{\textit{et al.}}: Bare Demo of IEEEtran.cls for Computer Society Journals}
%



\IEEEtitleabstractindextext{%
\begin{abstract}
Disaggregated memory leverages recent technology advances in high-density, byte-addressable non-volatile memory and high-performance interconnects to provide a large memory pool shared across multiple compute nodes. 
Due to higher memory density, memory errors may become more frequent. 
Unfortunately, tolerating memory errors through existing memory-error protection techniques becomes impractical due to increasing storage cost. 
This work proposes replication-aware memory-error protection to improve storage efficiency of protection in data-centric applications that already rely on memory replication for performance and availability.
It lets such applications lower protection storage cost by weakening the protection of each individual replica, but still realize a strong protection target by relying on the collective protection conferred by multiple replicas.
 
\end{abstract}

}

\maketitle

\IEEEdisplaynontitleabstractindextext

%
\IEEEpeerreviewmaketitle

\IEEEraisesectionheading{\section{Introduction}\label{sec:introduction}}

\begin{figure}[!t]
\centering
\includegraphics[width=3in]{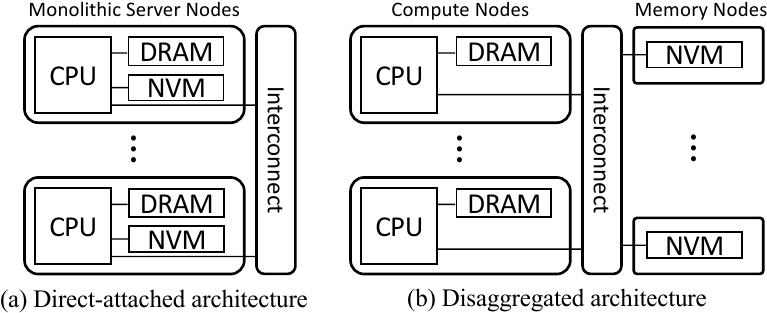}
\caption{Rack-scale non-volatile memory (NVM) architectures}
\label{fig:rack-scale-nvm}
\vspace{-0.5cm}
\end{figure}

\IEEEPARstart{R}{ecent} technology advances in high-density, byte-addressable non-volatile memory (NVM)\ignore{~\cite{wang:model-nvm:micro:2020}}
and low-latency interconnects\ignore{~\cite{knebel:genz:hotchips:2019,pinto:thymesis:micro:2020}}
have enabled building rack-scale systems with a large disaggregated memory pool shared across decentralized compute nodes~\cite{pinto:thymesis:micro:2020, tsai:dpm:atc:2020}.
\ignore{
Although memory disaggregation can be applied to both DRAM~\cite{pinto:thymesis:micro:2020} and NVM~\cite{tsai:dpm:atc:2020}, 
we focus on NVM disaggregation due to its potential for building memory pools of higher capacity. 
}
Unlike conventional rack-scale systems that are built with monolithic servers, each of which contains memory that is directly attached to the processor (Figure \ref{fig:rack-scale-nvm}(a)), disaggregated architectures decouple the processor from memory into separate compute and memory nodes  (Figure \ref{fig:rack-scale-nvm}(b)). 
This improves 
\ignore{
(i) separate evolution and scaling of processing and memory, which lets tailoring the compute-to-memory ratio to the specific needs of the workload, 
}
(i) memory utilization as memory is shared across multiple compute nodes, and (ii) failure isolation as compute and memory nodes can fail independently. 

A key challenge with building a large disaggregated memory pool based on NVM is efficiently tolerating memory errors at scale.
Nanoscale NVM technologies can be denser but also less reliable than DRAM due to their higher random raw bit error rate (RBER)~\cite{zhang:pm-chipkill:micro:2018}.
Thus, while adopting NVM technologies is desirable as it enables disaggregated memory to scale to petabyte-order capacities, it also makes memory errors more frequent~\cite{faraboschi:mem-centric-os:hotos:2015}.
Hence, efficient techniques for mitigating memory errors will be key to building cost efficient disaggregated memory systems.
Unfortunately, traditional mitigation techniques, such as parity and chipkill-correct, which can detect or correct memory errors, incur expensive storage costs.
Recent work has explored optimizations for chipkill-correct for NVM~\cite{zhang:pm-chipkill:micro:2018}, but storage cost remains  expensive ($\sim27\%$).

In this paper, we make the case for \emph{Replication-Aware Memory Protection} (\ramp).
\ramp seeks to further improve storage cost of memory error protection by leveraging the insight that many data-centric applications targeted by disaggregated memory~\cite{pinto:thymesis:micro:2020}, such as key-value stores and data analytics, replicate data to improve performance and availability.
Such applications will likely continue to replicate their data across multiple memory nodes in disaggregated memory to avoid a memory node from becoming a performance bottleneck or a single point of failure.
For example, a recent disaggregated key-value store~\cite{tsai:dpm:atc:2020} and a disaggregated operating system (OS)~\cite{shan:legoos:osdi:2018} replicate memory contents to improve availability.

\ramp lets applications employing replication to control the hardware-level memory error protection of individual replicas at each memory node to reduce storage overhead.
Instead of trying hard to prevent uncorrectable memory errors within a single memory node, an application can accept the possibility of uncorrectable errors, and employ weaker but lower-overhead resilience within individual memory nodes while relying on the collective protection conferred by the presence of available replicas in other memory nodes to tolerate a memory error.

To help applications determine the right protection strength of individual replicas, a key contribution of this paper is an analytical model that models the impact of individual protection strength on the collective protection strength.
We demonstrate the utility of our model by applying it to a recent chipkill-correct design~\cite{zhang:pm-chipkill:micro:2018}.
By weakening the chipkill protection of each individual replica, we reduce storage cost from $27\%$ down to $17.7\%$ while we attain the same protection level as the original design through the collective protection of multiple replicas. 
\vspace{-0.25cm}
\section{Disaggregated Memory}\label{sec:background}

\subsection{Enabling technologies and architecture}
Disaggregated memory builds on recent technology advances on two fronts.
First, non-volatile memory (NVM) technologies, such as phase-change memory (PCM), resistive random access memory (ReRAM), and commercially available Intel Optane DC Persistent Memory (DCPMM), provide byte-addressable persistent storage accessible via load/store instructions, rather than I/O requests. In addition to non-volatility, these technologies provide the potential for increased memory density and increased energy efficiency relative to DRAM. DCPMM has 2x higher read latency and 8x lower write bandwidth than DRAM, but it is up to 10x faster than Flash. 
Second, high-performance interconnects provide sub-microsecond access latency to remote memory~\cite{knebel:genz:hotchips:2019,pinto:thymesis:micro:2020}, while future interconnects based on silicon photonics are expected to further reduce latency~\cite{knebel:genz:hotchips:2019}.

These two technology advances provide the building blocks for constructing disaggregated memory architectures, where decentralized compute and memory nodes are interconnected by a high-performance system interconnect. 
Compute nodes mainly provide processing capability, but they also include a small amount of DRAM memory used as a local cache.
Memory nodes provide memory capacity in the form of NVM by attaching standard NVM subsystems to the network. 
Although the microarchitecture design of NVM subsystems is more complex than conventional DRAM subsystems\ignore{~\cite{wang:model-nvm:micro:2020}}, at a high level, NVM subsystems follow similar chip structure and system organization as DRAM subsystems: 
a memory controller is connected to memory modules through one or more channels, and each module provides an interface for accessing data stored across multiple chips, with chips comprising arrays of NVM bit cells. 

\subsection{Memory failures}
\label{sec:failure-model}
Disaggregation provides separate fault domains between processing and memory, meaning that the failure of a compute node does not render disaggregated memory unavailable, and vice versa, that is when a memory node fails, compute and other memory nodes continue to function. 
In this work, we focus on memory node failures caused by memory errors.

Memory errors may occur due to a variety of reasons.
First, memory errors may occur due to NVM bit cell errors. 
Bit errors are random in nature and can be caused by permanent faults due to limited and variable endurance, and transient faults due to resistance drift and read disturb\ignore{~\cite{yoon:freep:hpca:2011}}. 
Raw bit error rate (RBER) in PCM and ReRAM is significantly higher than in DRAM and ranges from $10^{-3}$ to $10^{-5}$~\cite{zhang:pm-chipkill:micro:2018}, depending on the technology and time since last write or refresh.
Second, memory errors may also occur when other components of the memory subsystem fail. 
Since NVM subsystems follow similar organization as DRAM subsystems, NVM subsystems will likely suffer from similar failures, including memory controller and memory channel failures due to \ignore{transient failures due to signal disturbances, and permanent failures due to }
faults in logic and transmission circuitry. 

To protect against memory errors, NVM subsystems maintain error correcting codes (ECC) computed over data. 
These codes can detect and correct a small number of errors.
For example, single error correction double error detection (SEC-DEC) uses parity to detect up to two-bit errors or correct a single-bit error.
Chipkill uses wider ECC to protect against multi-bit errors and chip failures.
Detectable but uncorrectable memory errors (DUE), which are detected but cannot be corrected by ECC, can cause memory node failures.
Non-detectable memory errors (NDE), which are non-detected and potentially miscorrected by ECC, do not cause memory node failures but may cause silent data corruption (SDC), which is also higly undesirable. 
For dense NVM with high RBER, simply extending existing memory protection mechanisms with stronger codes to achieve a low uncorrectable bit error rate (UBER) and low SDC rate incurs prohibitive storage overheads~\cite{zhang:pm-chipkill:micro:2018}.
These overheads remain significant ($\sim27\%$) despite recent efforts on improving storage
efficiency~\cite{zhang:pm-chipkill:micro:2018}. 

\vspace{-0.25cm}
\section{Replication-Aware Memory-error Protection}
\label{sec:shepherd}

We propose a flexible software-defined protection architecture called \emph{\textbf{R}eplication-\textbf{A}ware \textbf{M}emory-error \textbf{P}rotection} (\ramp)
to efficiently tolerate memory errors due to random bit cell errors in disaggregated memory.
We focus on random bit cell errors, as we expect these to represent the majority of memory errors because of the high RBER of high-density NVM.

\begin{figure}[!t]
\centering
\includegraphics[width=3in]{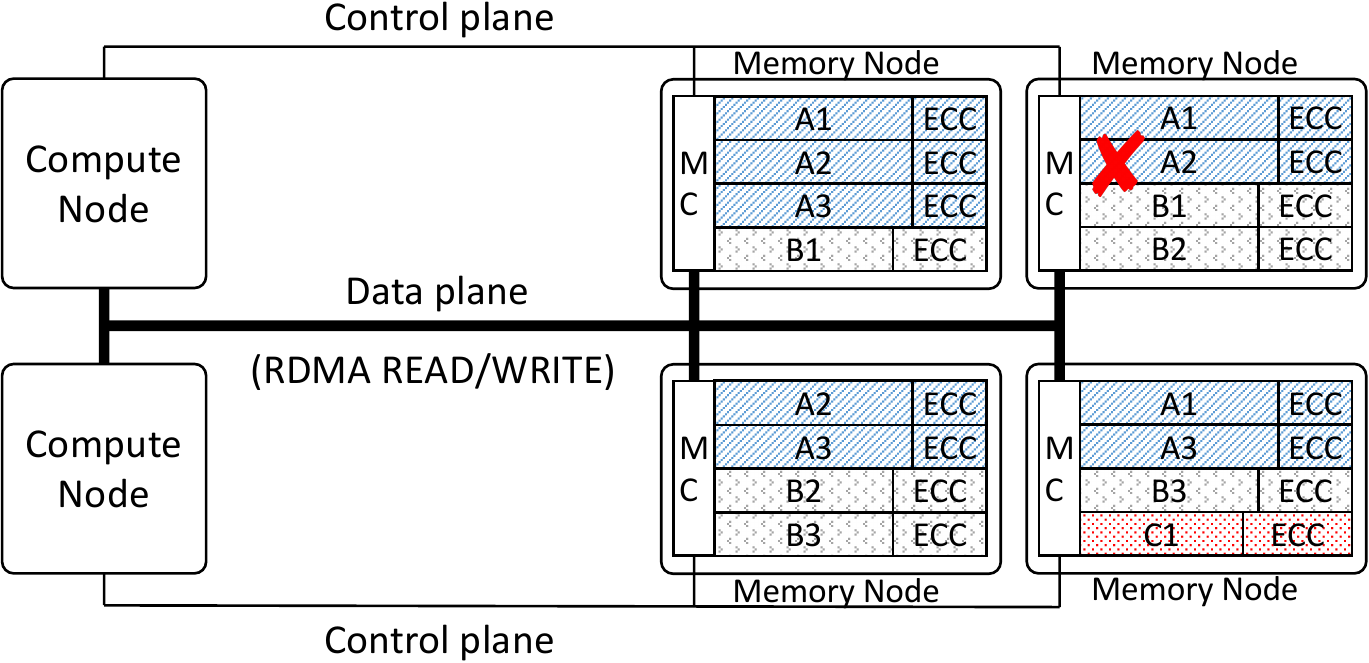}
\caption{\ramp architecture. }
\label{fig:ramp-architecture}
\vspace{-0.25cm}
\end{figure}


\ramp enables co-designing application-level replication together with hardware-level memory protection to improve storage efficiency. 
\ignore{Maintaining multiple replicas across memory nodes enables the power of many choices. Instead of trying hard to eliminate uncorrectable memory errors within a single memory node using strong but expensive codes, \ramp accepts the possibility of uncorrectable errors.}
Applications that maintain multiple replicas across memory nodes can employ weaker but lower-storage-overhead ECC within individual replicas.
While weaker ECC increases failure rate of individual replicas,  
applications can rely on the multiple choices offered by available replicas in other memory nodes to correct a memory error. 
\ignore{Similarly, applications can rely on checksums to detect silent corruptions that may evade the weaker ECC.}
For example, Figure~\ref{fig:ramp-architecture} shows three applications A, B, and C with different degrees of replication and levels of protection.
Application A maintains three replicas per data item so it uses weak ECC, relying on the collective protection of multiple replicas to tolerate the increased per-replica error rate.
\revisionhighlight{Application C uses no replication so it deploys strong ECC as it relies exclusively on ECC to tolerate memory errors.}

\subsection{Architecture}

Figure~\ref{fig:ramp-architecture} shows \ramp's system architecture. 
Compute nodes communicate with memory nodes through a control plane and a data plane.
Each memory node exposes its local NVM by attaching its memory controller (MC) to the control and data plane via a network interface card (not shown).
An application running on a compute node uses the control plane to dynamically configure hardware-level memory error protection at the memory nodes to provide a target UBER and SDC rate.
\revisionhighlight{
The application configures memory protection method (\eg, SEC-DEC, chipkill) and code strength at the granularity of individual pages.
\ramp supports page-granularity protection by augmenting the virtual memory page table and TLB to include protection information for each page. 
\ignore{
similarly to Virtualized ECC~\cite{yoon:virtualized-ecc:asplos:2010}.
}
\ignore{
This allows software to tune hardware ECC protection strength, trading off storage overhead, hardware UBER and SDC rate. 
Configuration capability can range from letting software select a memory protection scheme from a fixed set of protection schemes (\eg, SEC-DEC, chipkill) to providing complete flexibility in choosing the protection method and level~\cite{yoon:virtualized-ecc:asplos:2010}.
}
After configuration, the application uses the data plane to access data stored in the memory nodes using fast one-sided remote DMA (RDMA) reads and writes.
The memory controller processing the RDMA request uses the page protection information to identify which protection technique to use for any given memory access.
}

\ignore{
    Memory nodes expose any uncorrectable memory errors to compute nodes.
    Compute nodes tolerate such errors using software-level replication.
}

Memory nodes report DUEs to compute nodes for further handling. 
\ignore{
In contrast to previous work~\cite{shan:legoos:osdi:2018} where a memory node becomes unavailable upon a DUE, memory nodes expose DUEs to compute nodes for further handling. 
}
\ignore{
In contrast to previous work where uncorrectable memory errors crash a memory node, thus making all replicas stored on that node unavailable, our fine-grained failure model enables the replication protocol to continue using replicas stored in other memory regions unaffected by the memory error.
}
\ignore{
    The memory controller uses hardware ECC to detect and correct memory errors.
    The controller transparently corrects correctable errors, and silently returns invalid data for undetectable errors.
    For detectable but uncorrectable errors, the controller raises a hardware exception in response to uncorrectable memory errors. 
}
After reporting, a memory node remains operational and continues to serve memory accesses to non-failed memory regions, thus improving availability.
Memory nodes leverage existing hardware error reporting mechanisms, such as Intel Machine Check Architecture (MCA), to report DUEs. 
\ignore{
    For error reporting mechanisms, that do not provide a mechanism for detecting store failures, like Intel Machine Check Architecture (MCA), the NIC issues an additional read after a write to check success of the write.
}
\ignore{
    A lightweight service processor on the memory node handles the exception and returns an error as a response to the RDMA request by piggybacking on the existing error reporting mechanism of RDMA.
}
\ignore{
    To serve control plane operations and support fine-grain error reporting, the memory nodes include a lightweight service processor.
}

\subsection{Tolerating memory errors through replication}

\ramp targets applications that already employ replication for performance and availability, enabling them to leverage replication to correct DUE errors efficiently.
\ramp leaves the implementation of the replication method to the application for flexibility, dictating only some minimal requirements.
\ignore{
    Because targeted applications already employ replication for performance and availability (\cref{sec:ramp:idea}), we do not expect this requirement to significantly burden applications. 
}

For each replicated data item, an application maintains multiple replicas across memory nodes.
Applications map each replica to a memory node and memory region, and configure the hardware protection strength of each replica to meet a target UBER and SDC rate.
Applications may track and blacklist failed memory regions to avoid mapping replicas to regions with known errors.
When an application trying to access a data item faces a DUE, it attempts to correct the memory error using another replica. 
\ignore{
On read failure, it redirects the request to another replica.
On write failure, if the memory node remains operational, then it may attempt to write to another memory region within the same memory node. Otherwise it the memory node fails completely, software issues the writes to another memory node, and also remaps/migrates (asynchronously) all other replicas of the failed memory node. 
}

\revisionhighlight{
Applications can implement any block-level static homogeneous replication method, including primary-backup replication, chain replication, quorum-based replication, and MDS erasure coding. Static requires a fixed number of replicas whose protection strength does not change dynamically, thus relieving \ramp from having to support frequent protection changes. Homogeneous requires all replicas to have the same protection strength, thus simplifying replica strength reasoning.
}

\subsection{Choosing replica protection strength}

A key challenge in applying \ramp is choosing the right hardware protection strength of individual replicas. 
Weakening hardware-level protection of individual replicas lowers storage cost but makes DUEs and NDEs more frequent, increasing UBER and SDC rates respectively.
We can recoup the lost UBER by correcting a DUE using available replicas, at the expense of a performance overhead to access and process additional replicas.
However, we cannot always rely on replicas to recoup the lost SDC rate. 
This is because a NDE that silently corrupts data may not trigger replication-based correction, unless the application can detect the error through other means, such as checksumming~\cite{zhang:pangolin:atc:2019}.
Hence, SDC rate may limit how much we can weaken individual replica strength.

\ignore{
Lowering protection strength, increases UBER and SDC of individual replicas.
Using available replicas to correct a memory error recoups the lost UBER.
It doesn't recoup the lost SDC because a silent error that goes undetected does not trigger a correction using replication.
SDC rates are typically lower than UBER rates, so this gives some space for improvement.
So, we can lower protection strength, as long as combined UBER of replicas and SDC of individual replicas remain above a target threshold.

This in turn makes accessing additional replicas because of a memory error more frequent, which introduces extra processing and communication and increases performance overhead.
}

To help application and system designers choose protection strength, we develop an analytical model that estimates the expected reliability and expected performance overhead when using available replicas to correct a DUE. 
For the reliability, we estimate the combined DUE resulting from using replicas to correct a memory error. 
Because NDE does not benefit from replication, we do not compute a combined NDE. 
However, we do estimate the NDE of each individual replica as a lower reliability bound. 
For the performance overhead, we compute the average number of additional replicas that are read to correct a memory error.
We assume that reading and processing each replica contributes fixed network bandwidth and CPU overhead per replica.

Our analytical model targets block-level replication, which is a common replication approach. 
The model differentiates between logical and physical blocks.
The logical block is the unit of recovery, that is the smallest unit of data that can be recovered by the replication protocol.
To enable recovery, a replication protocol maps a logical block to multiple physical block replicas stored across multiple memory nodes.
Reading a logical block may involve reading one or more physical blocks. 

\ignore{
Since we focus on protection techniques against random bit cell errors, our model focuses on read failures caused by uncorrectable memory errors due to random bit cell errors. 
Our model ignores other correlated failures that affect multiple bits and blocks, such as chip or channel failures due to logic circuit errors.
}

\ignore{
The model uses the following parameters: CPU cache-line size: $c$:, Physical-block size: $b$, Cache-line failure probability due to DUE: $p_c$, Physical-block failure probability due to DUE: $p_b$.
}

\begin{table}
\caption{Analytical model symbol notation}
\label{tab:model}
\vspace{-0.2cm}
\centering
\begin{tabular}{lp{6.5cm}}
\textbf{Symbol(s)} & \textbf{Description} \\
$c$, $b$        & \scriptsize{Cache-line size and physical-block size}\\
$\pcdue$, $\pcnde$  & \scriptsize{Cache-line failure probability due to DUE and NDE}\\
$\pbdue$, $\pbnde$  & \scriptsize{Physical-block failure probability due to DUE and NDE}\\
$\pldue$  & \scriptsize{Logical-block failure probability due to DUE}\\
\end{tabular}
\vspace{-0.25cm}
\end{table}

The model uses the symbol notation shown in Table \ref{tab:model}.
Cache-line failure probability is the probability to fail when reading a cache-line worth of data from the memory system.
This probability depends on the memory protection scheme and the RBER of the NVM technology.
Although the RBER in NVM increases with the amount of time since last write or refresh, 
to simplify our model, we use a single worst-case value that is based on the RBER at the end of the refresh period\ignore{, which can range from a week to a year}. 
Physical-block failure probability is the probability to fail when reading a physical-block worth of data from the memory system.
Successfully reading a physical block entails successfully reading all the cache lines that comprise the block. 
We can derive the physical-block failure probability as follows:
\ignore{from the cache-line failure probability:}

\vspace{-0.2cm}
\begin{equation*}
\pbdue= 1 - (1-\pcdue)^{b/c} \quad \mathrm{and} \quad \pbnde= 1 - (1-\pcnde)^{b/c}
\end{equation*}

We next study two common application-level redundancy schemes: primary-backup replication and erasure coding.

\mypar{Primary-backup replication}
For each logical block, the replication protocol maintains a primary physical block and a sequence of N-1 backup physical replica blocks.
When a compute node needs to read a logical block, it first reads the primary physical block. If the read fails because of a DUE, then it tries the next backup physical block in the sequence, continuing this process until it successfully reads a block. 
When the compute node exhausts trying all available physical blocks without successfully reading one, the logical-block read fails with an uncorrectable memory error.

The probability to have a DUE when reading a logical block is the joint probability of all physical-block reads to fail:
\[
\pldue = \pbdue^{N}
\]

\ignore{
\fixme{The probability to have a NDE when reading a logical block is the union probability of all physical-block reads to fail}:

\[
\plnde = \sum_{i=0}^{N-1} (1-\pbdue)^i \pbnde
\]
}

The average number of additional physical blocks that are read after failing to read the first physical block is:

\[
\begin{split}
a_r&= -1 + \sum_{i=0}^{N-1} \pbdue^i(1-\pbdue)(i+1)
\end{split}
\]


\mypar{Erasure coding}
Erasure coding provides redundancy without the overhead of complete replication.
Erasure coding divides a logical block into K physical blocks and recodes them into N physical replica blocks, where N $>$ K.
When a compute node needs to read a logical block, it can perform the read using any K physical blocks of the N physical blocks. If the compute node fails to read any of the physical blocks (due to an uncorrectable memory error), then it tries another physical block. 
When the compute node exhausts trying all available physical blocks without successfully reading K blocks, the logical-block read fails with an uncorrectable memory error.

The probability to have a DUE when reading a logical block is the probability to have at least N-K+1 physical blocks fail due to a DUE:

\begin{equation*}
\pldue = \sum_{i=N-K+1}^{N}\binom{N}{i}\pbdue^{i}(1-\pbdue)^{N-i}
\end{equation*}

\ignore{
    \begin{equation*}
    \pldue = \sum_{i=N-K+1}^{N}f(i,N,\pbdue)
    \end{equation*}
    
    \begin{equation*}
    \mathrm{where}\quad
    f(k,n,p) = \binom{n}{k}p^{k}(1-p)^{n-k}
    \end{equation*}
    
    The probability to have a NDE is the number of ways in which we can read K physical blocks plus extra physical blocks due to DUE multiplied by the probability to have physical blocks fail due to DUE multiplied by the probability to have at least one physical block fail due to a NDE: 
    
    \begin{equation*}
    \plnde = \sum_{i=0}^{N-K}\binom{N}{K+i}f(i,K+i-1,\pbdue)\sum_{j=1}^{K-1}f(j,K-1,\pbnde)
    \end{equation*}
}

The average number of additional physical blocks that are read after failing to read any of the first K physical blocks is:

\begin{equation*}
a_r= -K + \sum_{i=0}^{N-K} \binom{N}{K+i} \binom{K+i-1}{i}\pbdue^{i}(1-\pbdue)^{K-1}(K+i)
\end{equation*}

\noindent where each sum term is the number of ways in which we can read physical blocks multiplied by the probability to have physical blocks fail due to DUE multiplied by the number of physical blocks read.

\begin{figure}[tb]
\centering
\includegraphics[width=3.5in]{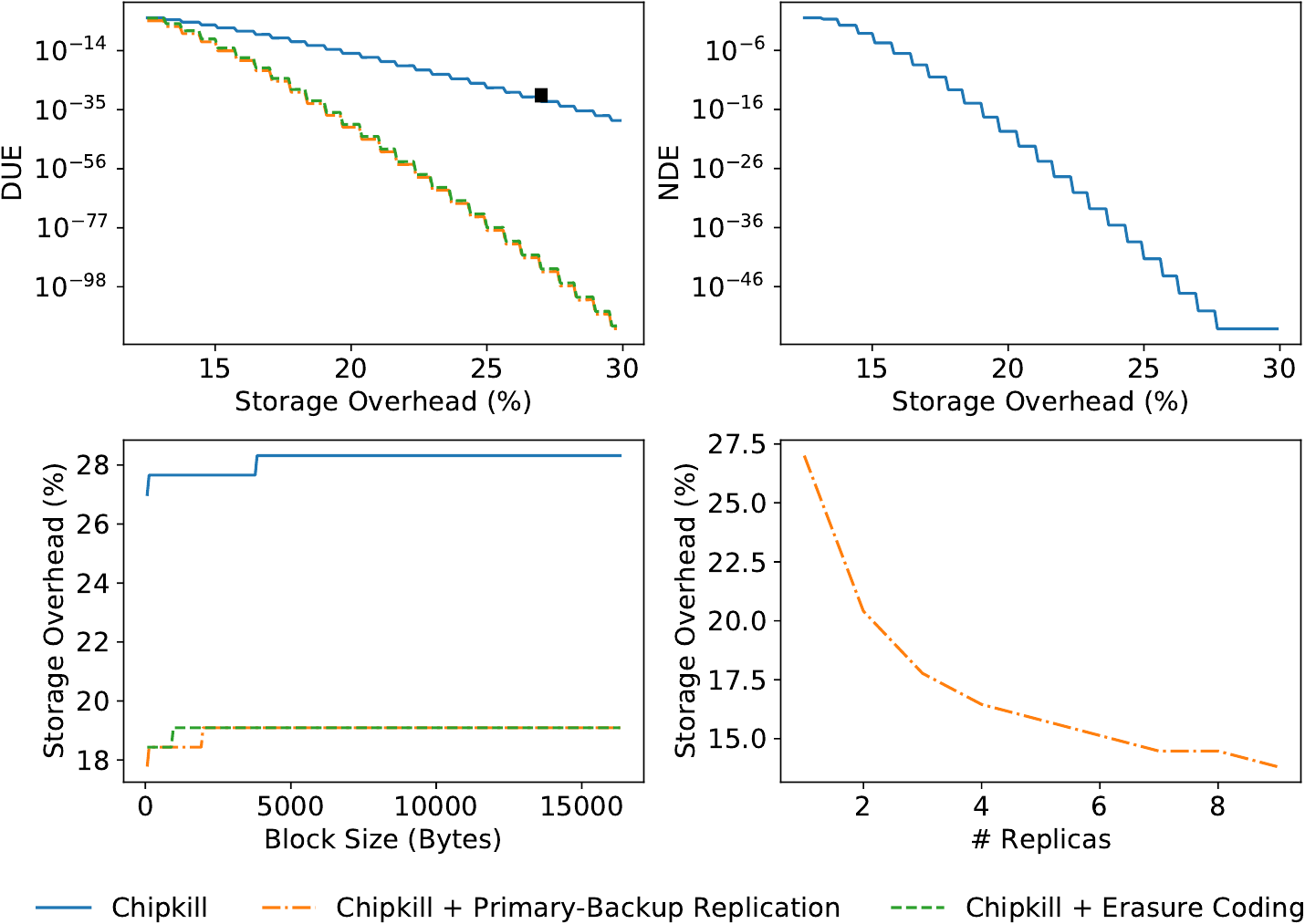}
\caption{DUE, NDE, and storage overhead for different chipkill protection schemes.
The solid rectangle (in the top-left figure) marks the DUE and storage overhead of the original chipkill design~\cite{zhang:pm-chipkill:micro:2018}.
NDE is shown only for baseline chipkill as it is independent of replication and identical for all chipkill schemes.
}
\label{fig:overhead}
\vspace{-0.4cm}
\end{figure}

\vspace{-0.25cm}
\section{Replication-Aware Chipkill-Correct}

We use our analytical model to study trade offs between reliability and storage overhead for a recent chipkill design~\cite{zhang:pm-chipkill:micro:2018}. 
The design achieves storage efficiency through a two-tier protection scheme: 
(i) a performance tier reuses the chip failure protection bits to opportunistically correct bit errors at high performance, and 
(ii) a storage-optimized tier uses long ECC codewords to correct at low storage cost bit errors that are detected but uncorrected by the performance tier.
The storage-optimized tier uses a BCH(2312,2048,22) code for each ECC codeword. BCH(n,k,t) uses a codeword of length $n=k+t(\left \lceil{log_2 (k)} \right \rceil+1)$ to correct $t$ bad bits when protecting $k$ bits of data~\cite{zhang:pm-chipkill:micro:2018}.

We show how we use our model to further optimize the storage-optimized tier. First, we need to compute the base cache-line failure probabilities of the storage-optimized tier.
We compute the cache-line failure probability due to DUE as the 
the probability that the storage-optimized tier fails to correct multiple bit errors in the BCH codeword (which happens when there are at least $t$ bit errors):

\begin{equation*}
\pcdue= \sum_{i=t+1}^{n}\binom{n}{i}{RBER}^i\cdot{RBER}^{n-i}
\end{equation*}

\ignore{
We compute the cache-line failure probability due to DUE as the product of two terms:
the probability that the performance tier fails to correct a bit error (whose value equals to 0.018 as is taken from the original design~\cite{zhang:pm-chipkill:micro:2018}) and 
the probability that the storage-optimized tier fails to correct multiple bit errors in the BCH codeword (which happens when there are at least $t$ bit errors):

\begin{equation*}
\pcdue= 0.018 \times \sum_{i=t+1}^{n}\binom{n}{i}{RBER}^i\cdot{RBER}^{n-i}
\end{equation*}
}

\noindent We compute the cache-line failure probability due to NDE following the analysis of Kim and Lee~
\cite{kim:undetected-error-bch:ieee-tc:1996}.
We assume an NVM technology with $RBER=2\times10^{-4}$, as in~\cite{zhang:pm-chipkill:micro:2018}.

We then use our model to estimate the combined DUE rate resulting from using available replicas to correct DUEs. 
We study three protection schemes: a baseline scheme that relies solely on chipkill (without redundancy) to protect blocks and two application-level redundancy schemes.
For the two redundancy schemes, we choose parameters so that they can tolerate up to two replica failures (following standard practice),
that is N=3 for primary-backup replication and N=5 and K=3 for erasure coding.
For all three schemes, we vary storage overhead by varying the strength of the BCH code used by the storage-optimized tier to protect individual replica blocks.
\revisionhighlight{
We vary strength by varying the number of $t$ bit errors that can be corrected by the BCH code.
}
We assume uniform access to all logical blocks and that all physical blocks are equally vulnerable to memory errors.
\revisionhighlight{
All replicas use the same ECC.
}

Figure~\ref{fig:overhead} plots combined DUE rate and NDE rate of individual physical blocks as a function of storage overhead.
For each replication scheme, the storage overhead is calculated over a corresponding baseline that employs the same replication scheme but without chipkill protection. 
For the top two figures, we use a physical block size equal to the cache line size, that is 64 bytes.
For the bottom-left figure, we vary the block size and plot the storage overhead sustained to achieve the same level of DUE as the original chipkill design.
\revisionhighlight{
For the bottom-right figure, we vary the number of replicas and plot the storage overhead sustained to achieve the same level of DUE as the original chipkill design.
}

We observe that both primary-backup replication and erasure coding can achieve the same level of DUE as the original ckipkill design ($\sim10^{-33}$ DUE rate), albeit at about $9\%$ less overhead.
\revisionhighlight{
For a target SDC rate of $10^{-22}$~\cite{zhang:pm-chipkill:micro:2018}, we need to provision an extra $2.4\%$ overhead, bringing the storage overhead savings down to $6.6\%$.
}
Although not shown, the relative performance overhead is negligible, less than $10^{-11}$.
Moreover, we observe diminishing returns in storage-overhead-savings as we increase the number of replicas, suggesting that \ramp could be more beneficial with low replication factors. Overall, these results confirm our main hypothesis: by weakening the protection of each individual replica, we can lower the storage overhead while we can rely on the combined protection conferred by multiple replicas to meet a stronger protection target.

\section{Conclusion}
We presented \ramp, our early work on efficiently tolerating memory errors in disaggregated memory systems based on high-density NVM.
\ramp gives applications (that employ replication for availability and performance) the flexibility to relax the protection strength of memory protection.
Our preliminary results show that such flexibility can bring notable savings in storage cost without sacrificing overall protection.

\vspace{-0.1cm}
\ifCLASSOPTIONcompsoc
  \section*{Acknowledgments}
\else
  \section*{Acknowledgment}
\fi

We sincerely thank Yanos Sazeides, Stavros Volos and the anonymous reviewers for their feedback on earlier
versions of this manuscript. This project has received funding from the European Union’s Horizon 2020 research and innovation programme under the Marie Skłodowska-Curie grant agreement No 101029391.

\vspace{-0.2cm}
\bibliographystyle{plain}
\interlinepenalty=10000
\bibliography{main}  

\end{document}